\documentclass{article}
\usepackage{spconf,amsmath,graphicx}

\usepackage[hidelinks]{hyperref}
\usepackage[table,dvipsnames]{xcolor}
\usepackage{booktabs}
\usepackage[nohyperlinks,nolist]{acronym}
\usepackage{amsmath,amssymb,amsfonts}
\usepackage[capitalize,nameinlink]{cleveref}
\usepackage{csquotes}
\usepackage{microtype}
\usepackage{pifont}
\usepackage{flushend}
\usepackage[moderate]{savetrees}
\usepackage[backend=biber, bibencoding=utf8,
            style=ieee,
            maxbibnames=10,
            doi=false,
            url=false,
            isbn=false]{biblatex}
\AtEveryBibitem{%
  \ifentrytype{inproceedings}{%
    \clearfield{pages}%
    \clearlist{publisher}%
    \clearlist{organization}%
    \clearlist{location}}{}%
  \ifentrytype{misc}{%
    \clearfield{eprintclass}{} }}

\begin{acronym}
\acro{DFT}{discrete Fourier transform}
\acro{FM}{flow matching}
\acro{TTS}{text-to-speech}
\acro{ODE}{ordinary differential equation}
\acro{SDE}{stochastic differential equation}
\acro{RTF}{real-time factor}
\acro{WER}{word error rate}
\acro{MCD}{mel-cepstral distance}
\acro{LSD}{log-spectral distance}
\end{acronym}

\newcommand{\B}[1]{\textbf{#1}}

\definecolor{darkred}{rgb}{0.55, 0.0, 0.0}

\newcommand{\U}[1]{\underline{#1}}
\newcommand{\xhat}{\hat{X}}

\definecolor{uhhblue}{RGB}{0,156,209}
\definecolor{uhhgreen}{RGB}{66, 178, 60}
\definecolor{uhhred}{RGB}{226,0,26}
\definecolor{uhhblack}{RGB}{0,0,0}
\definecolor{uhhstone}{RGB}{59,81,91}
\newcommand{\cmark}{\color{uhhgreen}{\ding{52}}}
\newcommand{\xmark}{\color{uhhred}{\ding{54}}}

\title{REAL-TIME STREAMING MEL VOCODING WITH GENERATIVE FLOW MATCHING}
\name{Simon Welker$^{\star}$\thanks{We acknowledge funding by the German Federal Ministry of Research, Technology and Space (BMFTR) under grant agreement No. 01IS24072A (COMFORT).} \qquad Tal Peer$^{\star}$ \qquad Timo Gerkmann$^{\star}$}
\address{$^{\star}$Signal Processing Group, Department of Informatics, University of Hamburg, Hamburg, Germany}

\begin{document}
\maketitle
\begin{abstract}
The task of Mel vocoding, i.e., the inversion of a Mel magnitude spectrogram to an audio waveform, is still a key component in many text-to-speech (TTS) systems today. Based on generative flow matching, our prior work on generative STFT phase retrieval (DiffPhase), and the pseudoinverse operator of the Mel filterbank, we develop \emph{MelFlow}, a streaming-capable generative Mel vocoder for speech sampled at 16 kHz with an algorithmic latency of only 32 ms and a total latency of 48 ms. We show real-time streaming capability at this latency not only in theory, but in practice on a consumer laptop GPU. Furthermore, we show that our model achieves substantially better PESQ and SI-SDR values compared to well-established not streaming-capable baselines for Mel vocoding including HiFi-GAN. %
\end{abstract}
\section{Introduction}
\label{sec:intro}

Mel vocoding, also known as Mel spectrogram inversion, is the task of converting a magnitude-only Mel spectrogram back into a waveform. It has long been a cornerstone of various speech processing tasks, especially for text-to-speech methods, which often use a two-stage approach involving a Mel vocoder as the second stage \cite{shen2018natural,kim2020glowtts,popov2021gradtts}. We are interested here in a \emph{streamable} variant of Mel vocoding with a fixed total latency $\Lambda$, where the Mel spectrogram is converted into a waveform frame by frame. Streamable vocoding is an important building block for natural real-time conversations with a \ac{TTS} model. Unlike the typical turn-based communication style, where full utterances are provided and processed by the user and the model in turn, streamable operation allows for an interaction that is much more human-like in nature and not constrained by the inherent waiting associated with non-streaming systems.

A few works have previously investigated streamable vocoding \cite{mustafa2021streamwise,mustafa2023framewise}, but to our knowledge none have explored the use of diffusion-based models for streamable Mel vocoding.
In FreeV \cite{lv2024freev}, the authors explore non-streaming Mel vocoding, and propose to use the Mel filterbank's Moore-Penrose pseudoinverse in order to initialize a neural vocoder. The authors show that this greatly aids model efficiency while improving resynthesis quality. We use these ideas to extend our prior work on diffusion-based STFT phase retrieval, DiffPhase \cite{peer2023diffphase}, to Mel vocoding.

Our contributions here are as follows: \B{(1)} we combine our prior work on diffusion-based STFT phase retrieval \cite{peer2023diffphase} with the main idea of FreeV \cite{lv2024freev} and a generative interpolating flow matching approach \cite{welker2025flowdec} to realize streamable generative Mel vocoding; \B{(2)} we develop a custom frame-wise causal generative DNN and iterative inference scheme with a total latency of $\Lambda = \text{48 ms}$ not only when batch-processing a Mel spectrogram, but in real-world streaming inference on a consumer laptop GPU; \B{(3)} we show that our model achieves substantially better PESQ and SI-SDR than state-of-the-art baseline models including HiFi-GAN \cite{kong2020hifigan} while staying competitive in non-intrusive metrics; \B{(4)} we provide, to our knowledge, the first public code repository and model checkpoint for streamable Mel vocoding\footnote{Code and checkpoints will be provided after acceptance.}.

\section{Background}
\label{sec:methods}
In the following, we introduce diffusion- and flow-based speech processing and relate them to streaming inference. Since the two are closely connected, we will call both \emph{diffusion-based}.
As fundamental definitions for streaming inference in STFT-based processing, we define the \emph{algorithmic latency} of a method as the STFT window duration, and the \emph{total latency} as the STFT window duration plus one STFT frame shift, subject to the constraint that the model that processes each frame must complete its operation within the duration of a single STFT frame shift. We define the \ac{RTF} of a streaming model as the processing time for a single frame divided by the frame shift duration, and say that a model is streaming-capable if it can process frame-by-frame and achieves \ac{RTF} $< 1$ in practice.

\subsection{Diffusion- and flow-based speech processing}
Diffusion-based speech enhancement and dereverberation was introduced by SGMSE \cite{welker2022speech} and SGMSE+ \cite{richter2023speech}, which define an interpolating diffusion process from corrupted speech spectra $Y \in \mathbb{R}^{T \times F}$ to clean speech spectra $X \in \mathbb{R}^{T \times F}$ based on a \ac{SDE}, where $T$ is the number of signal frames and $F$ is the frame dimensionality. SGMSE+ \cite{richter2023speech} showed high speech naturalness and robustness to data shifts and has since led to a large collection of downstream works. DiffPhase \cite{peer2023diffphase} builds upon SGMSE+ and frames STFT phase retrieval as a signal enhancement task, considering the loss of phase information as an auxiliary corruption process. Since diffusion-based methods usually require many DNN calls for inference ($N > 10$), FlowDec \cite{welker2025flowdec} proposes to formulate an interpolating flow matching process instead, which empirically works well with $N < 10$ for the complex task of neural codec post-filtering. We follow FlowDec here and define the corruption process as an \ac{ODE} interpolating continuously from $X_1 := X$ to $X_0 := Y + \sigma_y \varepsilon$, where $\varepsilon$ is an i.i.d. standard Gaussian noise vector of the same shape as $Y$, and $\sigma_y$ is a scalar hyperparameter. We refer to \cite{welker2025flowdec} for the full detailed derivation. We will choose $Y$ as a Mel-corrupted phaseless STFT spectrogram, which we detail in \cref{sec:streaming-mel-vocoding}.

\vspace{-.7em}
\subsection{Streaming Diffusion Inference}
In diffusion-based signal enhancement, we use a learned DNN $f_\theta$ with parameters $\theta$ and a numerical \ac{ODE} or \ac{SDE} solver to process a degraded input sequence $Y$ into an enhanced sequence $\xhat \approx X$. Within these solvers, we call $f_\theta$ for $N$ times in a row, starting from a noisy sequence $Y_0 = Y+\varepsilon$ with an independent Gaussian noise sample $\varepsilon \in \mathbb{R}^{T \times F}$. We assume that $f_\theta$ has a finite receptive field of size $R$ and is frame-causal, i.e., there exists no $t$ such that $\xhat[t]$ depends on any input frame $Y[t+n], n>0$. For ease of illustration, we use the equidistant Euler scheme for \acp{ODE} with $N$ steps in the following and assume the model $f_\theta$ was trained with an interpolating flow matching objective \cite{welker2025flowdec}. This inference process generates intermediate partially-denoised sequences $Y_1, Y_2, \ldots, Y_{N-1} \in \mathbb{R}^{T \times F}$ and finally the fully denoised sequence $Y_N$, which each depend on their respective prior sequences as follows:
\vspace{0.3em}
\begin{equation}
\begin{split}
    Y_1[t] &= Y_0[t] + \Delta\tau \cdot f_\theta(Y_0[t-R, \ldots, t], \Delta\tau), \label{eq:diff-solver}\\
    Y_2[t] &= Y_1[t] + \Delta\tau \cdot f_\theta(Y_1[t-R, \ldots, t], 2\Delta\tau), \\ %
    \scalebox{0.6}{$\vdots$} \\  %
    Y_N[t]   &= Y_{N-1}[t] + \Delta\tau \cdot  f_\theta(Y_{N-1}[t-R, \ldots, t], (N-1)\Delta\tau),\hspace{-.5em}
\end{split}
\end{equation}
where $\tau \in [0,1]$ is the continuous diffusion time and $\Delta\tau = \frac{1}{N}$ is the discretized diffusion time-step. After completing this process, $Y_N$ is fully denoised and we can treat it as the clean sequence estimate $\xhat:=Y_N$. Recursively collapsed, the overall procedure has an effective receptive field size of $(N \cdot R)$.

Given this, it may seem that efficient streaming diffusion inference with a large ($>10\,\text{M}$ parameters) DNN $f_\theta$ is not viable in a real-time setting: to generate a single output frame $\xhat[t] \in \mathbb{R}^F$ we must run the entire diffusion process backwards, doing so we must process all $(NR-1)$ past frames plus the newest frame, and we cannot exploit time-wise parallelism for efficiency since frames come in one-by-one.

One approach to circumvent this problem is the Diffusion Buffer \cite{lay2025diffusion}, which proposes to couple diffusion time and physical time in a fixed-size buffer, i.e., more noise is added to newer frames in the buffer. The buffer is then progressively denoised, sliding to the right by one frame at a time and outputting the oldest now fully denoised frame. This realizes a computationally efficient scheme at a fixed algorithmic latency equal to the buffer duration. A key downside is that the algorithmic latency, while sub-second, is still relatively large. The authors found that a buffer with fewer than 20 frames did not yield satisfying quality, i.e., one finds an effective minimum algorithmic latency of $(20+1) \cdot 16\,\text{ms} = 340\,\text{ms}$ with an STFT hop size of 16 ms.

\begin{figure}[t]
    \centering
    \includegraphics[width=0.92\linewidth]{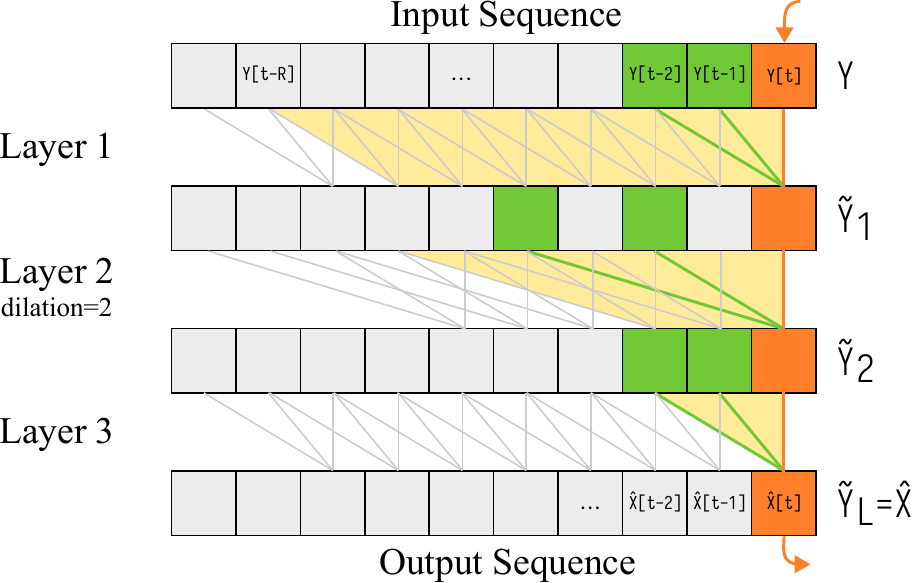}
    \vspace{-.5em}
    \caption{The inference scheme for a single new frame (orange) in a simplified frame-causal DNN. While the corresponding output frame has a receptive field size (yellow) of 9 in the input, only three frames must be evaluated in every layer since all required past results (green) are already available.}
    \label{fig:streaming-cnn-concept}
\end{figure}

\section{Proposed Methods}
\begin{table*}[t]
    \centering
    \vspace{-6pt}  %
    \caption{Mean metric values for the Mel vocoding task.
    \emph{S} indicates streaming-capable methods.
    \emph{HiFi-GAN (SB)} is the 16 kHz LibriTTS model published by SpeechBrain \cite{ravanelli2024speechbrain}, while \emph{HiFi-GAN (U)} is the official 22.05 kHz Universal-V1 model \cite{kong2020hifigan}.
    MelFlow with $N=25$ steps is not streaming-capable on our GPU and serves as a baseline for achievable quality.
    \B{Best}, \U{Second best}.
    }
    \vspace{3pt}
    \label{tab:melflow-results}
    \resizebox{0.98\textwidth}{!}{
    \begin{tabular}{llrrrrrrrrrl}
        \toprule
        \textbf{\textsc{Dataset}} / Method & {\small PESQ$\uparrow$} & {\small ESTOI$\uparrow$} & {\small SISDR$\uparrow$} & {\small DistillMOS$\uparrow$} & {\small WVMOS$\uparrow$} & {\small WER$\downarrow$} & {\small NISQA$\uparrow$} & {\small LSD$\downarrow$} & {\small MCD$\downarrow$} & {\small S} \\
        \midrule
        \multicolumn{10}{l}{\small\textbf{\textsc{EARS-WHAM v2 (16 kHz)}}}\\
        $M^\dagger$ + RTISI-DM \cite{peer2024flexible} & 2.86 & 0.88 & -29.1 & 2.66 & 2.04 & 7.5\% & 2.82 & 0.91 & 2.95 & \cmark \\
        HiFi-GAN (SB) \cite{kong2020hifigan,ravanelli2024speechbrain} & 2.99 & \U{0.90} & -29.9 & 4.21 & 3.02 & \U{7.3\%} & 3.91 & \U{0.77} & \U{2.41} & \xmark \\
        MelFlow \B{(ours)} & \U{4.12} & \B{0.96} & \B{-8.8} & \U{4.32} & \U{3.15} & \B{7.2\%} & \B{4.14} & 1.00 & 3.28 & \cmark \\
        MelFlow $N=25$ \B{(ours)} & \B{4.25} & \B{0.96} & \U{-10.6} & \B{4.34} & \B{3.28} & \B{7.2}\% & \U{4.07} & \B{0.70} & \B{1.70} & \cmark \\
        \midrule
        \midrule
        \multicolumn{10}{l}{\small\textbf{\textsc{LibriTTS (16 kHz)}}}\\
        $M^\dagger$ + RTISI-DM \cite{peer2024flexible} & 2.67 & 0.89 & \U{-25.6} & 2.56 & 2.64 & 5.4\% & 2.84 & \U{0.96} & \U{2.90} & \cmark \\
        HiFi-GAN (SB) \cite{kong2020hifigan,ravanelli2024speechbrain} & \U{3.03} & \U{0.92} & -25.8 & \U{4.02} & \B{3.86} & \U{5.1}\% & \U{4.06} & \B{0.87} & \B{2.09} & \xmark \\
        MelFlow \B{(ours)} & \B{3.97} & \B{0.95} & \B{-14.5} & \B{4.08} & \U{3.67} & \B{3.5\%} & \B{4.32} & 1.12 & \U{2.90} & \cmark \\
        \midrule
        \multicolumn{10}{l}{\small\textbf{\textsc{LibriTTS at higher sampling rates (evaluated at 16 kHz)}}}\\
        HiFi-GAN (U) \cite{kong2020hifigan} {\footnotesize (22.05 kHz)} & \U{2.99} & \U{0.92} & -26.4 & \U{4.08} & \U{3.95} & \B{3.2\%} & \U{4.11} & \U{0.75} & \U{1.91} & \xmark \\
        Vocos \cite{siuzdak2024vocos} {\footnotesize (24 kHz)} & \B{3.67} & \B{0.96} & \B{-24.7} & \B{4.11} & \B{4.00} & \U{3.9\%} & \B{4.13} & \B{0.67} & \B{1.22} & \xmark \\
        Parallel WaveGAN \cite{yamamoto2020parallelwavegan} {\footnotesize (24 kHz)} & 2.83 & 0.90 & \U{-25.8} & 3.71 & 3.77 & 5.7\% & 3.75 & 0.78 & 2.70 & \xmark \\
        StyleMelGAN \cite{mustafa2021stylemelgan} {\footnotesize (24 kHz)} & 2.70 & 0.89 & -26.3 & 3.86 & 3.78 & 6.4\% & 3.92 & 0.83 & 3.06 & \xmark \\
        \bottomrule
    \end{tabular}
    }
    \vspace{-6pt}
\end{table*}

\subsection{Streaming Mel Vocoding}\label{sec:streaming-mel-vocoding}
We extend the STFT phase retrieval method DiffPhase \cite{peer2023diffphase} to Mel vocoding through a change in the corruption model based on the ideas presented in FreeV \cite{lv2024freev}. DiffPhase solves the task of STFT phase retrieval by defining an interpolating diffusion process, with the starting point mean $Y$ being the phaseless magnitude spectrogram $Y = |X|+0j$ embedded into the complex plane, and the clean complex STFT coefficients $X$ as the target, where $j$ denotes the imaginary unit. 
Here, we instead define an interpolating flow \cite{welker2025flowdec} and propose to use a magnitude spectrogram subjected to a per-frame lossy Mel compression and pseudoinverse decompression as $Y$:
\vspace{.5em}
\begin{equation}
    Y[t] := \left|M^\dagger \left(M \big|X[t]\big|\right)\right| + 0j\vspace{.5em}
\end{equation}
where $M \in \mathbb{R}^{F_{\text{mel}} \times F_{\text{STFT}}}$ is the Mel matrix mapping an STFT frame to a Mel frame, $M^\dagger$ is its Moore-Penrose pseudoinverse, and $|X[t]|$ denotes the single magnitude spectrogram frame at time $t$.
This extends the DiffPhase task from only generating STFT phases to also enhancing STFT magnitudes.
Note that $(M|X[t]|)$ corresponds to the Mel frame at time $t$, and serves as the input Mel frame during inference. Since $M^\dagger$ may lead to negative values that are not meaningful for a magnitude STFT, we take the magnitudes again after $M^\dagger$, as also done in FreeV \cite{lv2024freev}. We follow the Mel configuration of HiFi-GAN \cite{kong2020hifigan} in the 16 kHz LibriTTS variant from SpeechBrain \cite{ravanelli2024speechbrain}, but use a 32 ms (512-point) Hann window instead of a 64 ms window for the STFT to reduce the algorithmic latency, keeping the 16 ms hop duration (50\% frame overlap). This scheme is streaming-capable as long as the enhancement model is streaming-capable, since $M^\dagger$ is a cheap per-frame operation.

\subsection{Efficient Streaming Diffusion Inference}
As an alternative to the Diffusion Buffer \cite{lay2025diffusion} that does not modify the diffusion process, we design a scheme that can perform streaming frame-wise inference while incurring no additional algorithmic latency and avoiding redundant computations. We build upon prior works \cite{hedegaard2022cins,mustafa2021streamwise} but formalize the methodology in detail to clarify the extension to diffusion/flow model inference. We first assume that $f_\theta$ is a causal convolutional neural network (CCNN) containing $L$ stacked causal convolution layers $\mathcal{C}_\ell$ each with stride 1, dilation $d_\ell$, and a kernel $K_\ell \in \mathbb{R}^{c_{o,\ell} \times c_{i,\ell} \times k_\ell}$ of size $k_\ell$, input channels $c_{i,\ell}$, and output channels $c_{o,\ell}$. We ignore bias terms here for simplicity. The DNN may also contain any operations that are point-wise in time (e.g., activations such as SiLU, and specific types of normalization layers). Each convolution layer has receptive field size $R_\ell = (k_\ell -1)d_\ell + 1$.
The first key observation here is that a causal convolution of kernel size $k$ and dilation $d$ depends on exactly $k$ frames of its input sequence: $Y[t-(k-1)d, t-(k-2)d, \ldots, t]$, and immediately when these input frames are available, its output $X[t]$ is fixed. Even when stacking multiple convolutions, the past outputs of layer $\mathcal{C}_{\ell-1}$ never change and do not need to be recomputed before being fed to layer $\mathcal{C}_{\ell}$. This lets us perform local caching throughout the DNN, where for every layer $\mathcal{C}_{\ell}$ we keep a rolling buffer $B_\ell \in \mathbb{R}^{c_{i,\ell} \times (R_\ell-1)}$ containing only $R_\ell-1$ past input frames.

When we receive a new input frame $y[t]$ to the DNN after previously having seen the input frames $y[\ldots, t-1]$, we only need to evaluate each layer's kernel $K_\ell$ at the latest time index $t$ on the buffer $B_\ell$ concatenated with the newest frame, producing a single output frame per layer, see \cref{fig:streaming-cnn-concept}:%
\vspace{.4em}%
\addtolength{\belowdisplayskip}{.1em}%
\begin{align}
    \tilde{Y}_{\ell=1}[t] &= \phi_1(K_1 \star Y[\underbrace{t-R_1, \ldots, t-1}_{=B_1}, t]), \\
    \tilde{Y}_{\ell=2}[t] &= \phi_2(K_2 \star \tilde{Y}_{\ell=1}[t-R_2, \ldots, t-1, t]),\quad\ldots, \\
    \tilde{Y}_{\ell=L}[t] &= \phi_L(K_L \star \tilde{Y}_{\ell=L-1}[t-R_L, \ldots, t-1, t]),
\end{align}%
\addtolength{\belowdisplayskip}{-.1em}%
where the $\tilde{Y}_\ell$ represent the intermediate activations, $\star$ denotes the convolution evaluated at only a single frame yielding an output in $\mathbb{R}^{c_{o,\ell} \times 1}$, and $\phi_\ell$ represents any operations between convolutions that are point-wise in time.
We can then set $\xhat[t] = \tilde{Y}_{\ell=L}[t]$ as the output frame of this single DNN call.
A prior work on \emph{Continual Inference Networks} (CINs) \cite{hedegaard2022cins} explores streamable implementations of various neural network components, where the authors note that strided convolutions incur a delay to their downstream modules and hence incur algorithmic latency, so we avoid their use here along the time dimension.

Our novel second key observation is that this scheme can be straightforwardly extended to diffusion/flow model inference with multiple DNN calls by tracking $N$ independent collections of cache buffers, one for each DNN call, resulting in $(N \cdot L)$ cache buffers in total. We denote this collection of buffers as $\mathbf B$ where $\mathbf B_{n,\ell} \in \mathbb{R}^{k_\ell \times F}$, $n=0,\ldots,N-1$ and $\ell=1,\ldots,L$. Starting from $Y_0[t]$, we can then process:
\vspace{.2em}
\addtolength{\belowdisplayskip}{.5em}
\begin{gather*}
    \left.
    \begin{aligned}
        \tilde{Y}_{1,\ell=1}[t] &= \phi_1(K_1 \star [\mathbf B_{1,1}, Y_0[t]]),\quad\ldots, \\
        Y_1[t] := \tilde{Y}_{1,\ell=L}[t] &= \phi_L(K_L \star [\mathbf B_{1,L}, \tilde{Y}_{1,\ell=L-1}[t]]),\\
    \end{aligned}
    \right\} \text{ Call 1 }\\[.5em]
    \left.
    \begin{aligned}
        \tilde{Y}_{2,\ell=1}[t] &= \phi_1(K_1 \star [\mathbf B_{2,1}, Y_1[t]]),\quad\ldots, \\
        Y_2[t] := \tilde{Y}_{2,\ell=L}[t] &= \phi_L(K_L \star [\mathbf B_{2,L}, \tilde{Y}_{2,\ell=L-1}[t]]),
    \end{aligned}
    \right\} \text{ Call 2 }\\
    \scalebox{0.6}{$\vdots$}\hspace{22em}\\
    Y_N[t] := \tilde{Y}_{N,\ell=L}[t] = \phi_L(K_L \star [\mathbf B_{N,L}, \tilde{Y}_{N,\ell=L-1}[t]]),\hspace{2.5em}
\end{gather*}
where $[\cdot, \cdot]$ denotes concatenation along the time dimension and we again treat $\xhat := Y_N$ as the enhanced output sequence. We conceptually include the ODE solver steps, \cref{eq:diff-solver}, in $\phi_L$. After every operation, the respective buffer $\mathbf B_{n,\ell}$ is shifted to the left by one frame and the newest input is inserted on the right. The buffers are all initialized with zeros, matching the offline case since our convolutions use zero-padding.
\addtolength{\belowdisplayskip}{-.5em}

This overall scheme now performs the same number of computations as the original offline model, spread across physical time. Importantly, it also yields exactly the same result (up to numerical floating-point inaccuracies) as feeding the entire sequence to the causal model in an offline fashion. This enables efficient parallel training, and the use of the trained weights for streaming inference without any train/test mismatch.

\subsection{DNN architecture}
We build a custom streaming-capable frame-wise causal DNN as a modification of the NCSN++ \cite{song2021scorebased} U-Net architecture often used for diffusion-based speech enhancement \cite{richter2023speech}. Concretely, we
\B{(1)} replace all convolutions with causal convolutions through zero-padding;
\B{(2)} perform down-and upsampling only along frequency and never along time, and instead use a dilation of 2 along time where there was a time-wise downsampling;
\B{(3)} remove all attention layers,
\B{(4)} replace the noncausal GroupNorm with a sub-band grouped BatchNorm inspired by \cite{chang2021subspectral}, with 4 groups along frequency and the original NCSN++ grouping along channels;
\B{(5)} remove extra skip connections along the upsampling path and always employ addition instead of concatenation for feature fusion;
\B{(6)} use two ResNet blocks per level and three levels of downsampling;
\B{(7)} do not use a weight exponential moving average.
Our network has 27.9 M parameters.  %

\section{Experiments}
\label{sec:experiments}

\subsection{Model training}
We train our model with an interpolating flow matching objective as in \cite{welker2025flowdec}, on the clean utterances from the EARS-WHAM v2 dataset\footnote{\url{https://github.com/sp-uhh/ears_benchmark}} \cite{richter2024ears} containing around 87 hours of clean speech, which we downsample from 48 to 16 kHz. We use 2-second random crops, four GPUs and a batch size of 12, and train for 200 epochs ($\sim$ 140k steps) with the SOAP optimizer \cite{vyas2025soap} using a decaying cosine learning rate schedule from $\lambda = 5 \cdot 10^{-4}$ to $\lambda = 10^{-6}$ and a linear warmup for the first 1000 steps. We set the flow matching noise level hyperparameter \cite{welker2025flowdec} to $\sigma_y = 0.25$. We compress STFT magnitudes with an exponent of $\alpha=0.5$ as in \cite{welker2022speech,richter2023speech}, but in contrast to these works we use an orthonormal FFT and do not perform any additional scaling.

\vspace{-.5em}
\subsection{Evaluation}
For evaluation we use the clean test set utterances of EARS-WHAM v2 \cite{richter2023speech} and the LibriTTS clean test set \cite{zen2019libritts}, both downsampled to 16 kHz. We run our MelFlow model at $N=5$ inference steps which is real-time capable on a consumer laptop GPU, but also compare with $N=25$ inference steps \emph{MelFlow*} with $N=25$ inference steps for achievable quality without computational constraints. As intrusive metrics, we use (wideband) PESQ \cite{rix2001pesq}, ESTOI \cite{jensen2016estoi}, SI-SDR \cite{leroux2018sdr}, \ac{MCD} \cite{kubichek1993mcd} and \ac{LSD} \cite{rabiner1993fundamentals}, as well as the \ac{WER} using Whisper \cite{radford2023robust} for automatic speech recognition. As non-intrusive quality metrics, we use DistillMOS \cite{stahl2025distillmos}, WVMOS \cite{andreev2023hifi}, and NISQA \cite{mittag2021nisqa}.

As baselines we use HiFi-GAN \cite{kong2020hifigan} in the 16 kHz LibriTTS variant published by SpeechBrain \cite{ravanelli2024speechbrain} and in the Universal-V1 22.05 kHz variant by the authors \cite{kong2020hifigan}, as well as Vocos \cite{siuzdak2024vocos}, Parallel WaveGAN\footnote{\label{footnote:pwg}\url{https://github.com/kan-bayashi/ParallelWaveGAN}} \cite{yamamoto2020parallelwavegan} and StyleMelGAN\footnotemark[3] \cite{mustafa2021stylemelgan}, all trained on LibriTTS and operating at 24 kHz. For baselines operating at a higher sampling rate, we run inference on the LibriTTS clean test set downsampled to each respective sampling rate, and then downsample the outputs to 16 kHz for calculating all metrics.
We further devise a non-learned streaming baseline which we call \emph{$M^\dagger$ + RTISI-DM}, using a streaming-capable algorithm for STFT phase retrieval \cite{peer2024flexible}.
Here we first apply the Mel pseudoinverse $M^\dagger$ and the absolute value to retrieve approximate STFT magnitudes. We then run RTISI-DM without look-ahead \cite{peer2024flexible} for STFT phase retrieval, using the Difference Map (DM) algorithm \cite{elser2003phaseretrieval} with $\beta=1.75$ and 50 iterations per frame, where we line-searched the best $\beta \in [-2, 2]$.

\section{Results}
\label{sec:results}

\begin{figure}
    \centering
    \includegraphics[width=\linewidth]{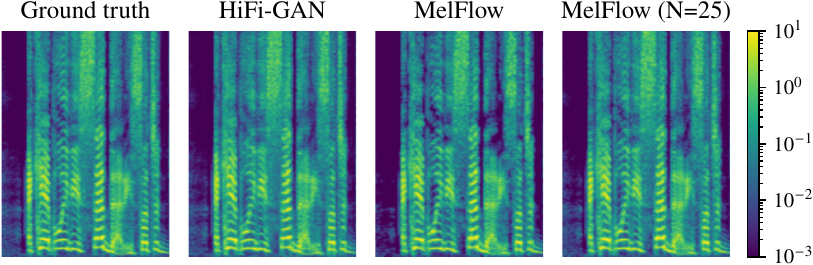}
    \caption{Resynthesized spectrograms for a 16 kHz EARS \cite{richter2024ears} snippet, comparing HiFi-GAN to MelFlow and MelFlow* (N=25, not streamable) and the ground truth. Note in particular the better recovery of high harmonics from MelFlow.}
    \label{fig:example-spectrograms}
    \vspace{-.5em}
\end{figure}

Our main results in terms of metrics are shown in \cref{tab:melflow-results}. We show that on EARS-WHAM v2, which our model was trained on, we outperform the 16 kHz SpeechBrain HiFi-GAN \cite{ravanelli2024speechbrain} in all metrics except LSD and MCD, though our streaming-incapable $N=25$ MelFlow variant outperforms HiFi-GAN in these metrics as well. For LibriTTS, we have similar findings, achieving the best PESQ and SI-SDR and performing better than, or similar to, all baselines in most metrics except WVMOS, LSD and MCD.
Regarding real-time streaming, on a \emph{NVIDIA RTX 4080 Laptop} GPU we find a processing time of $(N \times 2.71\,\text{ms})$ for $N$ inference steps, which allows for up to $N=5$ steps while staying within the budget of the 16 ms frame shift.
In \cref{fig:example-spectrograms}, we show an example resynthesized spectrogram from HiFi-GAN compared to MelFlow and MelFlow with $N=25$. It can be seen that MelFlow reconstructs high harmonics more faithfully than HiFi-GAN. While MelFlow at the default $N=5$ exhibits some loss of fine spectral details, this is not the case at $N=25$, which suggests a possibility to further improve MelFlow's output quality with few-step distillation techniques.

\section{Conclusion}
We have shown here that streaming Mel vocoding can be performed in real-time with MelFlow, a streaming generative flow matching model with a total latency of 48 ms, by designing an efficiently cached inference scheme and custom frame-causal DNN. While MelFlow is streaming-capable in contrast to the evaluated baseline methods, MelFlow nonetheless significantly outperforms several established Mel vocoder baselines including HiFi-GAN regarding PESQ and SI-SDR, and also performs competitively in non-intrusive quality metrics.

\clearpage
\renewcommand*{\bibfont}{\linespread{0.8}\selectfont\normalsize}
\section{References}%
\label{sec:refs}
\printbibliography[heading=none]

\end{document}